# Towards truly simultaneous PIXE and RBS analysis of layered objects in cultural heritage

Carlos Pascual-Izarra[1], Nuno P. Barradas[2], Miguel A. Reis[2], Chris Jeynes[3], Michel Menu[4], Bertrand Lavedrine[5], Jean Jacques Ezrati[4], Stefan Röhrs[4]

[1] Instituto de Estructura de la Materia, Consejo Superior de Investigaciones Científicas, C/Serrano 113 bis, Madrid E28006, Spain

[2] Instituto Tecnológico e Nuclear, E.N. 10, Sacavém Codex 2686-953, Portugal

[3] Surrey Centre for Research in Ion Beam Applications, School of Electronics, Computing and Mathematics, University of Surrey, Guildford GU2 7XH, United Kingdom

[4] Centre de recherche et de restauration des musées du France, CNRS UMR 171, Palais du Louvre - Porte des Lions 14, Quai François Mitterrand, Paris 75001, France

[5] Centre de recherches sur la conservation des documents graphiques, Muséum national d'histoire naturelle, 36, rue Geoffroy-Saint-Hilaire, Paris 75005, France

## Abstract

For a long time, RBS and PIXE techniques have been used in the field of cultural heritage. Although the complementarity of both techniques has long been acknowledged, its full potential has not been yet developed due to the lack of general purpose software tools for analysing the data from both techniques in a coherent way. In this work we provide an example of how the recent addition of PIXE to the set of techniques supported by the DataFurnace code can significantly change this situation. We present a case in which a non homogeneous sample (an oxidized metal from a photographic plate -heliography- made by Niépce in 1827) is analysed using RBS and PIXE in a straightforward and powerful way that can only be performed with a code that treats both techniques simultaneously as a part of one single and coherent analysis. The optimization capabilities of DataFurnace, allowed us to obtain the composition profiles for these samples in a very simple way.

**PACS:** 82.80.Ej; 82.80.Yc; 02.70.Uu; 07.05.Tp; 29.30.Kv; 41.75.Ak
**Keywords:** PIXE; X-rays; RBS; Rutherford backscattering; Fit; Cultural heritage; photography

# 1  Introduction

The Rutherford Backscattering Spectroscopy (RBS) and Particle Induced X-ray Emission (PIXE) techniques have been used for many years in the field of cultural heritage [1]. Furthermore, many groups combine them to gain a better insight on the composition of the samples due to the complementarity of both techniques as well to the fact that both spectra can be collected simultaneously [2,3].
But, while the experiments might be simultaneous, the analysis of the data has been traditionally done with separate software tools and different philosophies for each technique. Even when dealing with very simple samples, the RBS+PIXE analysis often required several iterations involving feedback from one technique into the other to obtain a final sample description that is consistent with both the PIXE and RBS spectra. In the case of not so simple samples (*e.g.*, in most cultural heritage problems where layered samples are involved), the analysis became too hard and reaching to a self-consistent solution implied a lot of effort and tailored experiment. These difficulties have prevented the full development of an otherwise ideal combination of techniques.

We have recently presented an innovative software tool (Data Furnace, incorporating the LibCPIXE module [4,5]) that is able to perform simultaneous and self consistent analyses of RBS and PIXE data (as well as of various other IBA techniques [6,7]). In this work we show an application of DataFurnace to the field of cultural heritage by analysing a photo plate (heliography) from J. N. Niépce dated to 1827. This photographic plate is made from a tin alloy and the image is formed by black alterations of the silver-shining plate. The black alterations are oxidation products, produced by a treatment of tin alloy by acids. In order to study the manufacturing process of this heliographic plate, a characterisation of the artificially corroded surface is needed.
The corrosion layer present in the photographic plate would be very difficult (if not impossible) to analyse correctly with conventional PIXE programs due to the presence of elements that change their concentration profile with depth, leading to an ill-posed problem beyond the capabilities of such codes. On the other hand, the approach used in this work takes advantage of two features of the DataFurnace tool: first, the analysis is not done one spectrum at a time, but rather the information from various spectra is considered in a truly simultaneous fit; secondly, advanced and robust fitting algorithms (such as Simulated Annealing) are available for fitting [6,7]. The analysis procedure is further discussed in section 3.

# 2  Experimental

RBS and PIXE spectra were obtained simultaneously at the micro-spot beam line of the AGLAE accelerator [1] using a 3MeV proton and 3MeV alpha beams at normal incidence angle (note: we use the term RBS even when non-Rutherford cross sections are involved [8]). The experiments were performed in a helium-rich atmosphere and this was taken into account for the analysis. The PIXE spectra were collected using a Si(Li) detector placed at $45°$ from the surface normal, with a $50\mu m$ Al filter foil and a $8\mu m$ Be window. The RBS spectra were collected with a surface barrier detector placed at $150°$ angle relative to the beam.

Two different spots of the same photographic plate were investigated for this work. The first one is a clean spot exposing the metallic compound layer of the original plate (mainly Sn and Pb) and the other is a dark spot showing corrosion. The dark spot was expected to be formed by oxides of the metals of the underlying layer.

To calibrate the X-ray detection efficiency of the system, spectra from a standard sample containing known amounts of Sn, Pb, Cu and various other elements were also collected. These experiments were also used to measure the solid angle and efficiency ratio between the PIXE and the RBS detectors.

# 3  Data Analysis

The information content of a single PIXE spectrum from the type of samples considered here is not enough to obtain a unique solution describing a concentration profile changing with depth. Since the traditionally used tools for analysing PIXE data only analyse one spectrum at a time, it is not possible to use them in a straightforward way. Similarly, analysing this kind of samples only with RBS also leads to ambiguous solutions. The traditional approach in these cases involves using some code to analyse the PIXE data, making assumptions like considering that the sample is homogeneous and obtaining a somewhat "averaged" composition of the sample, then using a RBS simulation program to guess a possible layered description of the sample compatible with the averaged values. Possibly, grazing-angle or low-energy PIXE spectra could also be used to gain information about near-surface regions and various iterations considering a number of different candidate composition profiles would be needed. In this process, the user would need to bounce between separate tools for RBS and PIXE analysis (although some degree of automation could be attained using an interface tool such as DAN32 [9]) until a solution satisfying all the experimental data is obtained. In contrast to that situation, the DataFurnace tool incorporating the LibCPIXE module does not perform separate fits for each spectrum but, rather, simultaneously fits all the experimental data provided about a given sample as a single block. Also, in comparison with the gradient based algorithms for optimization used in [10-14], DataFurnace can make use of a simulated annealing algorithm, which has been proved to be much more robust and less prone to stick to sub-optimal local solutions [15].

The procedure followed to analyse the spectra was similar to the one already published in [4] except in that some extra work was needed to deal with the fact that the fluence was not measured. First, the PIXE spectra were processed using QXAS [12] to subtract the background, identify the relevant X-ray lines and obtain their intensities, since DataFurnace works with these rather than with the raw PIXE spectra. Secondly, the RBS spectrum for 3MeV protons on the standard sample was used to determine the fluence during the 3MeV proton irradiation. This value, together with the PIXE data for the same irradiation, allowed us to obtain the efficiency curve for the Si(Li) detector. It is worth noting that the experimentally determined "efficiency" curve already incorporates a constant factor related to the unknown solid angle ratio between the RBS and the PIXE detectors.

Once the calibration was done, the clean spot of the photographic plate was analysed. For this, all the experimental data (namely, the proton PIXE data, as well as the proton and alpha RBS spectra) were fitted simultaneously using simulated annealing with DataFurnace. The free parameters during the fitting were: the sample composition (assuming that the clean plate was homogeneous in depth, besides a thin surface layer with O and C), the fluence for the different spectra (forcing it to be the same for the spectra acquired simultaneously) and the energy calibration parameters (within strict limits) in the case of the RBS spectra.

The analysis of the dark (corroded) spot was similar to that for the clean one except that the fit did not assume in-depth homogeneity. In order to reduce the number of free parameters in this fit, the bulk material present below the corrosion layer was assumed to have the same composition as that obtained from the clean spot. Similarly, the corroded layer was assumed to contain the same elements (but with free changing proportions) plus oxygen and carbon, as seen in the PIXE and RBS spectra. No further assumptions were made either in the thickness or concentration profiles describing the corrosion layer.

# 4 Results

Since the clean spot can be considered a homogeneous sample, the 3 MeV proton PIXE spectrum can be used with GUPIX [11] to obtain an independent characterisation of this sample in order to validate the DataFurnace results. The Table 1 shows a good agreement between the DataFurnace and GUPIX characterisations of the clean spot (note that for DataFurnace all the data was considered as opposed to just the single PIXE spectrum used for GUPIX). The good fit for the RBS spectrum of the same sample (shown in Fig. 1) further confirms the validity of the DataFurnace analysis.

Figure 2 shows the composition profiles for the elements present in the dark spot of the photographic plate. It can be seen that there is a relatively large amount of oxygen near the surface that gradually decreases with increasing depth. Conversely, the relative tin content is lower at the surface than in the bulk, while lead content is roughly constant all across the sample, suggesting that the corrosion is due to oxidation from the surface.

While manually fitting the data from the corroded sample is not an easy task, it is performed automatically in a few minutes by DataFurnace with a high degree of accuracy, as shown in Fig. 3. Note that the program used a total of 14 layers for describing the sample, the first being the He atmosphere and the last being the bulk non-corroded plate underlying the corrosion. The remaining layers describe the slowly varying concentration profiles in the corrosion layer.

# 5 Conclusions

We have given an example of how the gap that separated the software tools for PIXE and RBS data analysis has been filled with the recent incorporation of PIXE to the repertoire of techniques that can be analysed with the DataFurnace code. The combination of both techniques not only at the experiment level but also at the data analysis level shows great advantages as many problems can only be adequately solved if a coherent and automatic analysis is done. We have shown how three different data types (proton PIXE, proton RBS and alpha RBS) can be analysed simultaneously leading to a solution that satisfies all of them and that would not be obtained if we had followed a more traditional approach.

We find that the development of routine truly simultaneous PIXE and RBS analysis provides significant advantages to the ion beam analysis field in general, and to the cultural heritage studies in particular, as exemplified by the characterisation of the 1827's heliography layer shown here.
Finally, we want to emphasise that DataFurnace is not limited to simultaneous PIXE and RBS analysis. Multiple spectra from one or more of the supported techniques (PIXE, RBS, ERD and non-resonant NRA) can be arbitrarily combined and simultaneously fitted. For example, many PIXE spectra of the same sample acquired with different energy and/or geometry conditions can be used to analyse inhomogeneous samples.

# Bibliography

[1] Dran, JC Salomon, J Calligaro, T Walter, P. Ion beam analysis of art works: 14 years of use in the Louvre. *Nuclear Instruments and Methods in Physics Research Section B: Beam Interactions with Materials and Atoms* (2004) **219-20**: p. 15.

[2] Mando PA. Advantages and limitations of external beams in applications to arts & archeology, geology and environmental problems. *Nuclear Instruments and Methods in Physics Research Section B: Beam Interactions with Materials and Atoms* (1994) **85**: pp. 815-823.

[3] Mathis F, Moignard B, Pichon L, Dubreuil O & Salomon J. Coupled PIXE and RBS using a 6 MeV 4He2+ external beam: A new experimental device for particle detection and dose monitoring. *Nuclear Instruments and Methods in Physics Research Section B: Beam Interactions with Materials and Atoms* (2005) **240**: pp. 532-538.

[4] Pascual-Izarra C, Reis M & Barradas N. Simultaneous PIXE and RBS data analysis using Bayesian inference with the DataFurnace code. *Nuclear Instruments and Methods in Physics Research Section B: Beam Interactions with Materials and Atoms* (2006) **249**: pp. 780-783.

[5] Pascual-Izarra C, Barradas N & Reis M. LibCPIXE: A PIXE simulation open-source library for multilayered samples. *Nuclear Instruments and Methods in Physics Research Section B: Beam Interactions with Materials and Atoms* (2006) **249**: pp. 820-822.

[6] Barradas, NP Jeynes, C Webb, RP. Simulated annealing analysis of Rutherford backscattering data. *Applied Physics Letters* (1997) **71**: p. 293.

[7] Jeynes, C Barradas, NP Marriott, PK Boudreault, G Jenkin, M Wendler, E Webb, RP. Elemental thin film depth profiles by ion beam analysis using simulated annealing - a new tool. *Journal of Physics D-Applied Physics* (2003) **36**: p. R126.

[8] Amsel G. CUTBA (Cleaning Up the Tower of Babel of Acronyms) in IBA. *Nuclear Instruments and Methods in Physics Research Section B: Beam Interactions with Materials and Atoms* (1996) **118**: pp. 52-56.

[9] DAN32. http://www.microbeams.co.uk/.

[10] S. Fazinić. Intercomparison of PIXE Spectrometry Software Packages. International Atomic Energy Agency (IAEA). TECDOC 1342 (2003)

[11] Guelph Pixe Software Package (GUPIX). http://www.physics.uoguelph.ca/PIXE/gupix/.

[12] QXAS. http://www.iaea.org/OurWork/ST/NA/NAAL/pci/ins/xrf/pciXRFdown.php.

[13] RUMP. http://www.genplot.com/.

[14] M. Mayer. SIMNRA Users Guide. Max-Planck-Institut für Plasmaphysik. Technical Report IPP9/113 (1997)

[15] Emile Aarts JK. Simulated annealing and Boltzmann machines: a stochastic approach to combinatorial optimization and neural computing. Wiley, Chichester. 1989.

# Figures & Tables

| Element | Sn | Pb | Mn | Cu | As |
|---|---|---|---|---|---|
| at% (DataFurnace) | 96.24 | 2.20 | 0.62 | 0.48 | 0.32 |
| at% (GUPIX) | 95.81 | 2.01 | 1.43 | 0.47 | 0.27 |

*Table 1: Composition for the clean photographic plate, as obtained by DataFurnace using a simultaneous PIXE + H-RBS + He-RBS fit. The results obtained with GUPIX using just the 3 MeV H-PIXE spectra are also shown for comparison. Only elements present at more than 0.1 at% are tabulated.*

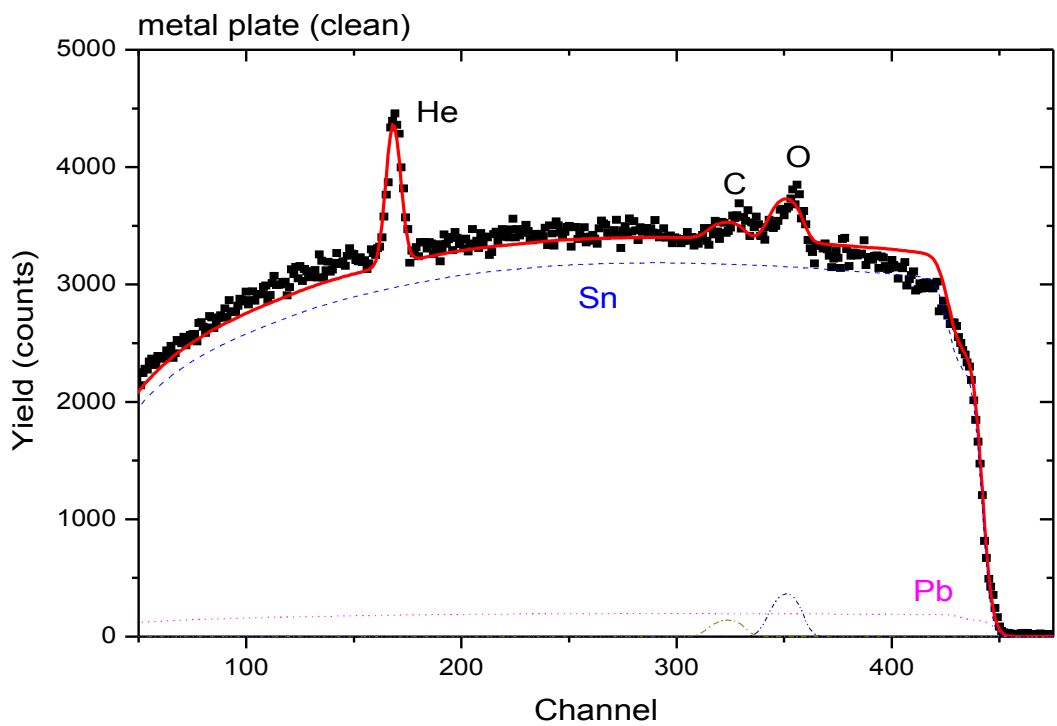

*Figure 1: Fitted RBS spectrum for 3MeV proton beam on the "clean" spot. Partial spectra for each element are also shown.*

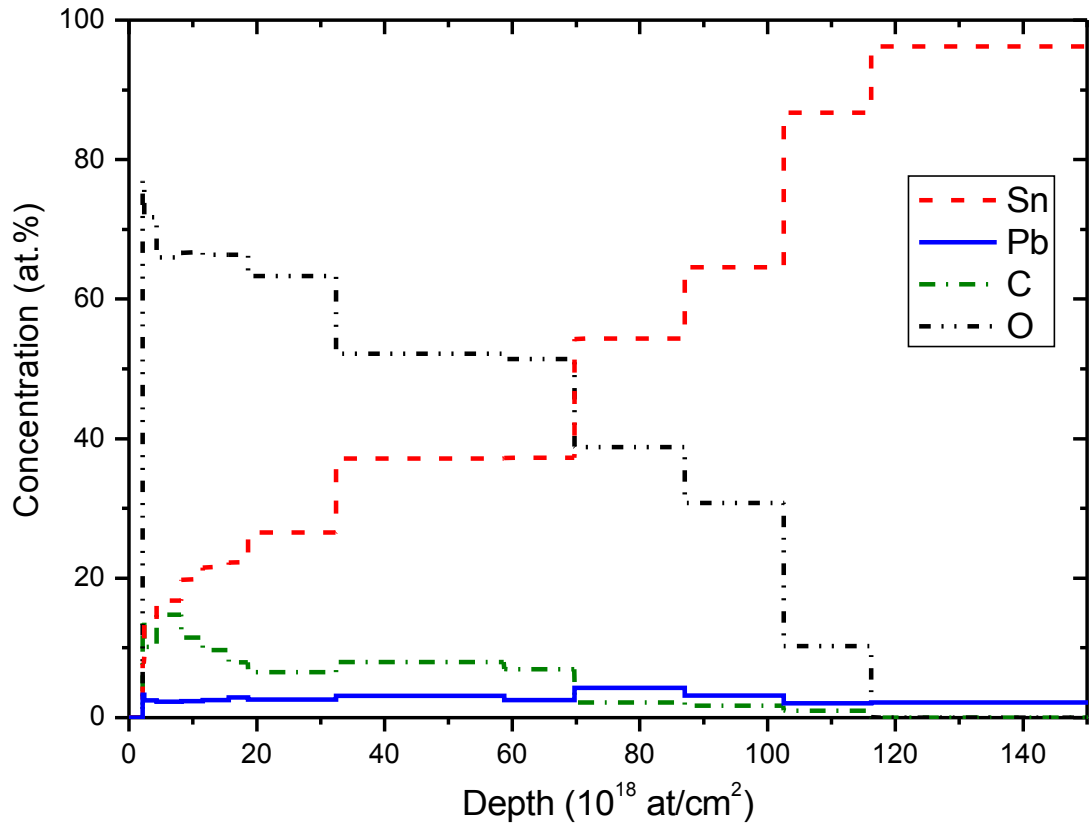

*Figure 2: Concentration profiles for the "dark" (corroded) area, as obtained by a simultaneous fit of data from PIXE, proton-RBS and alpha-RBS.*

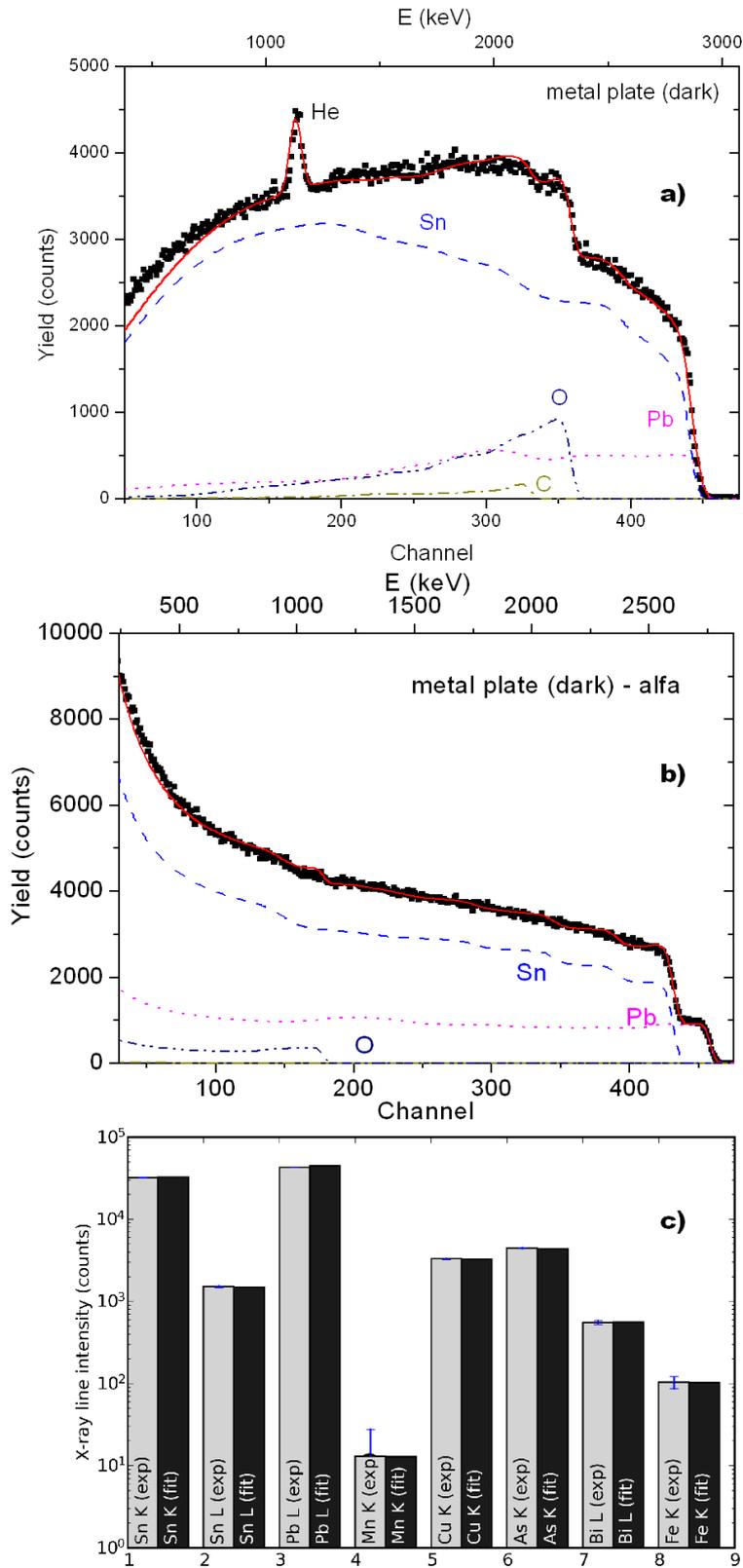

*Figure 3: Experimental data and fitting results for the "dark" (corroded) area. a) and b) show the RBS data with 3 MeV proton and alpha beams, respectively and c) shows the PIXE data for the most intense X-ray lines.*